\documentclass[12pt]{iopart}
\usepackage{amsfonts}
\begin{document}
\jl{1}
\title{Feynman Path Integral on the Noncommutative Plane.}

\author{Anais Smailagic\dag\footnote[1]{e-mail
        address: \texttt{anais@ictp.trieste.it}},
        Euro Spallucci\ddag\footnote[5]{e-mail
        address: \texttt{spallucci@trieste.infn.it}}
        }
        
        \address{\dag\ Sezione INFN di Trieste,\\
         Strada Costiera 11, 34014 Trieste,\\
         Italy}
        \address{\ddag\ Department of Theoretical Physics,\\
         University of Trieste, Strada Costiera 11, 34014 Trieste,\\
         Italy}

\date{\today}

\begin{abstract}
We formulate Feynman path integral on a non commutative plane using coherent 
states. The propagator for a free particle exhibits  UV cut-off induced by the 
parameter of non commutativity.
\end{abstract}


\maketitle
One of the difficulties when working with non commutative geometries is
to perform explicit calculations in terms of non commutative coordinates $x$. 
In non commutative quantum mechanics, it is possible to find a suitable combination 
of phase space coordinates which define new, {\it commuting} position coordinates, with non 
commutativity manifesting itself as an external magnetic field \cite{soliti}, \cite{noi}. 
  In quantum field theory similar approach cannot be applied since quantum fields are 
  functions of position coordinates  only, and momenta conjugate to these coordinates do 
  not appear. In this case it is customary to formulate a non commutative field 
  theory \cite{agm}  in terms of ordinary functions of commuting variables, endowed with 
  a ($\ast$-product) multiplication law  \cite{wwm}.
  However, in the formulation of non commutative theory based on $\ast$-product the basic 
  property of the non commutativity, i.e. the existence of a natural UV cutoff due
 to the uncertainty in position,  is not transparent. In particular,
 the free propagator is unaffected by the $\ast$-product as if non commutativity
 had no effect on it. Calculations are than performed using truncated series expansion in 
 the parameter characterizing the non commutativity of space. This approach leads to UV/IR 
 mixing phenomenon but one is still faced with 
 divergences to cure \cite{uvir} as in ordinary quantum field theory. It is possible that
 the complete resummation of the $\ast$-product  expansion would give a finite
 result but such a procedure is beyond  calculational capabilities (as is in ordinary 
 perturbation approach). For this reason the use of $\ast$-product, while perfectly 
 well-defined procedure, did not result in expected UV finiteness of non commutative field 
 theory. On the other hand, the existence of a minimal length in non commutative plane 
 should manifest itself already in the free propagator being the property of space (geometry)
  and not of interaction among fields.\\ 
 With hope to cure this flaw,  we have recently presented a new formulation of
  quantum field theory on the non commutative plane \cite{noilast}
which is based on coherent state formulation, instead of
$\ast$-product, and which is  explicitly UV finite.\\ 
In this report we would like to use the ideas introduced in
\cite{noilast} to formulate the Feynman path integral on the non commutative 
plane. The reason is that the Feynman path integral provides a common framework
both for quantum mechanics and field theory. As a an example of this connection
we shall calculate the Feynman propagator for a scalar particle as a sum over
paths in phase space. Comparing  the propagator found in this approach
with the one in \cite{noilast} we find complete agreement.\\

The core ingredient in the formulation of
non commutative models without $\ast$-product is the use of expectation values 
of operators between \textit{coherent states}. In this way, non commutativity
of coordinates is carried on by the Gaussian spread of coherent states.\\
Contrary to space coordinates, we choose to consider \textit{commutative} canonical
momenta since their non commutativity can be introduced through covariant
derivatives in the usual way.\\
 We start with non commutative two dimensional plane described by the set of 
 coordinates $\mathbf{q}_1$, $\mathbf{q}_2$ satisfying 

\begin{equation}
\left[\, \mathbf{q}_1\ , \mathbf{q}_2\, \right]=i\,\theta
\label{comm}
\end{equation}

The main difficulty in working with non commutative coordinates is that there
 are no common eigenstates of $\mathbf{q}_1$, $\mathbf{q}_2$ due to
 (\ref{comm}). Thus, one cannot work in coordinate representation
 of ordinary quantum mechanics. To bypass this problem
 we construct \textit{new} raising/lowering operators  in terms of coordinates only, as

\begin{eqnarray}
&&\mathbf{A}\equiv \mathbf{q}_1 + i\mathbf{q}_2\\
&&\mathbf{A}^\dagger\equiv \mathbf{q}_1 - i\mathbf{q}_2
\end{eqnarray}

 The coordinate analogue of the canonical commutator of creation and destruction operators 
 is
\begin{equation}
 \left[\, \mathbf{A}\ , \mathbf{A}^\dagger\, \right]=2\,\theta
\end{equation}

The \textit{Coherent States} \cite{glaub} are defined as eigenstates of the 
above operators in the following sense

\begin{eqnarray}
&& \mathbf{A}\, \vert\, \alpha\, \rangle =\alpha\,\vert\, \alpha\, \rangle \\
&& \langle\, \alpha\,\vert \, \mathbf{A}^\dagger =\,\langle\,
\alpha\,\vert\,\alpha^\ast
\end{eqnarray}
 
 The explicit form of the normalized coordinate coherent state is
 
  \begin{equation}
 \vert\, \alpha\, \rangle= \exp\left(-\frac{1}{2}\alpha\,\alpha^\ast\,\right)
 \exp\left(\,\alpha\, \mathbf{A}^\dagger \,\right)\, \,\vert\, 0\, \rangle
 \end{equation}
 
 We define \textit{expectation values} of non commuting 
 coordinates between coherent states as
 
 \begin{eqnarray}
  \langle\, \alpha\,\vert \,\mathbf{q}_1\, \vert\, \alpha\, \rangle &&=
 \langle\, \alpha\,\vert \, \frac{\mathbf{A}^\dagger + \mathbf{A}}{2} 
 \, \vert\, \alpha\, \rangle\nonumber\\
 &&=\frac{\alpha^\ast + \alpha}{2}=\mathrm{Re}\left(\,\alpha\,\right)\nonumber\\
 &&\equiv x_1
 \end{eqnarray}
 
 \begin{eqnarray}
  \langle\, \alpha\,\vert \,\mathbf{q}_2\, \vert\, \alpha\, \rangle &&=
 \langle\, \alpha\,\vert \, \frac{-\mathbf{A}^\dagger + \mathbf{A}}{2i} 
 \, \vert\, \alpha\, \rangle\nonumber\\
 &&=\frac{-\alpha^\ast + \alpha}{2i}=\mathrm{Im}\left(\,\alpha\,\right)\nonumber\\
 &&\equiv x_2
 \end{eqnarray}
 
 The vector $\vec x=\left(\, x_1\ , x_2\,\right)$ represents
 the \textit{mean position} of the particle over the non commutative plane.
 The advantage of the use of mean values is that they do not represent sharp 
 eigenvalues and thus can be measured simultaneously, in spite of 
 non commutativity of coordinates. In the same spirit, to any operator valued 
 function, $F\left(\, \mathbf{q}_1,\mathbf{q}_2\,\right)$ we  associate a 
 function $f(x_1\ , x_2)$. \\
 For future purposes, let us consider the non commutative version of the plane 
 wave operator $\exp i\,\left(\, \vec p\cdot\vec\mathbf{q}\,\right)$ where 
 $\vec p=\left(\, p_1\, p_2\,\right)$ is a real
 two-component vector. The mean value is calculated to be
 
 \begin{equation}
\langle\, \alpha\,\vert \, e^{\, i\, p_1\, \mathbf{q}_1 +i\,
    p_2\,\mathbf{q}_2\,}\, \vert\, \alpha\, \rangle =
\exp\left( -\theta\, \frac{\,\vec p\,{}^2}{4} + i\,\vec p\cdot \vec x\,\right)
\label{thetawave}
\end{equation}

  In order to obtain the above result we have exploited the Backer-Hausdorff 
  decomposition
 
 \begin{eqnarray}
  e^{\,i\, p_+\, \mathbf{A}^\dagger\, +i\, p_-\, \mathbf{A}\,}&&=
 e^{\,i\, p_+\, \mathbf{A}^\dagger}\, e^{\,i\, p_-\, \mathbf{A}}\, 
 e^{\,\frac{p_+\,p_-}{2}\,\left[\,\mathbf{A}^\dagger\ , \mathbf{A} \, \right]\,}
\nonumber\\
 &&=
 e^{\,i\, p_+\, \mathbf{A}^\dagger}\, e^{\,i\, p_-\, \mathbf{A}}\,
 e^{-\theta\,p_+\,p_-} \nonumber\\
 &&=
 e^{\,i\, p_+\, \mathbf{A}^\dagger}\, e^{\,i\, p_-\, \mathbf{A}}\,
 e^{-\theta\, \frac{\,\vec p\,{}^2}{4}}
 \end{eqnarray}
 
 where, we found it convenient to define
 
 \begin{eqnarray}
 && p_-\equiv \frac{ p_1 - i\, p_2}{2}\\
 && p_+\equiv \frac{ p_1 + i\, p_2}{2}
 \end{eqnarray}
 
 The mean value of the plane wave, which takes into account the
 non commutativity of coordinates, is  the key to the Fourier Transform
 of quantum fields in \cite{noilast}. At the same time, (\ref{thetawave}) 
 shows that the vector $\vec p $ is canonically conjugated to the
 \textit{mean position} $\vec x $ and can be given the meaning of
  \textit{mean linear momentum.} Thus, we can interpret (\ref{thetawave}) 
 as the wave function of a ''free point particle'' on the non commutative plane:
 
 \begin{equation}
 \psi_{\vec p}\left(\,\vec x \,\right)\equiv 
 \langle\, \vec p\,\vert \, \vec x\, \rangle_\theta \equiv \exp\left(-\theta\,
  \frac{\,\vec p\,{}^2}{4} + i\,\vec p\cdot \vec x\,\right)
  \label{onda}
 \end{equation}
 
  The amplitude between two states  of different mean position is the key ingredient
  in the formulation of the Feynman path integral. It is given by
 
 \begin{eqnarray}
 \langle\, \vec y\,\vert \, \vec x\, \rangle &&=
 \int \frac{d^2p}{(2\pi)^2}\,
 \langle\, \vec y\,\vert \, \vec p\,\rangle \langle \, \vec p\, \vert \,
  \vec x\, \rangle \nonumber\\
 &&=\left(\, \frac{1}{2\pi\theta}\,\right)\,
  \exp\left(-\frac{\left(\, \vec x -\vec y\,\right)^2}{2\, \theta}\,\right)
 \end{eqnarray}
 
 We thus get the important result that the effect of non commutativity
 in the scalar product between two mean positions is to substitute
 a Dirac delta function by a Gaussian whose
 half-width $\sqrt\theta$ corresponds to the minimum length
 attainable in fuzzy space. Now, we can proceed and formulate the Feynman
 path-integral for non commutative theories.\\
 One starts from the definition of  discretized transition amplitude
 between two nearby points \cite{feyhibbs}. With the help of (\ref{onda})
 we find 
 
 \begin{eqnarray}
\fl 
\langle\, \vec x_{(i+1)}\ , \epsilon\, \vert \, \vec x_{(i)}\ , 0\,\rangle
 &&=\langle\, \vec x_{(i+1)}\,  \vert\, e^{-i\epsilon H} \,\vert\,
  \vec x_{(i)}\,\rangle\nonumber\\
&&= \langle\, \vec x_{(i+1)}\,  \vert\,  1 -i\epsilon H  + O(\epsilon^2)
 \,\vert\,\vec x_{(i)}\,\rangle\nonumber\\
 &&= \int \frac{d^2p_{(i)}}{(2\pi)^2}\, \langle\, \vec x_{(i+1)}\,  \vert\, 
 \vec p_{(i)}
 \,\rangle \langle\, \vec p_{(i)}\, \vert \, \vec x_{(i)}\,\rangle
  e^{-i\epsilon H(\, \vec x_{(i)}\ ,\vec p_{(i)}\,) }\nonumber\\
 &&= \int \frac{d^2p_{(i)}}{(2\pi)^2} e^{i\left(\,\vec x_{(i+1)}- \vec x_{(i)}
 \,\right)\,
  \vec p_{(i)}}\exp\left(
  -i\epsilon H(\, \vec x_{(i)}\ ,\vec p_{(i)}\,)-\theta\,\vec
  p_{(i)}^2/2 \,\right)\nonumber\\
 \end{eqnarray}
 
 Following the usual steps of summation over discretized paths, and letting
 the number of intervals to infinity, one finds the non commutative version
 of the path integral for the propagation kernel
 
 \begin{equation}
 \fl 
K_\theta\left(\, x-y\ ;T\,\right)=N\int \left[\, Dx\,\right] \left[\,
  Dp\,\right]\exp\left\{\, i\int_y^x \vec p\cdot d\vec x -
  \int_0^T d\tau\left(\, H\left(\, \vec p\ ,\vec x\,\right) 
  + \frac{\theta}{2T}\,\vec p\,{}^2\right) 
  \,\right\} \label{kernel}
  \end{equation}
  
  The end result is that the path integral gets modified by the 
  Gaussian factor produced by non commutativity. 
  To emphasize the physical relevance of this modification let us calculate
  the propagator for the free  particle on the non commutative plane. Starting with 
  (\ref{kernel}) and performing the  integration over coordinates, as in
  \cite{euroj}, we find
  
  \begin{eqnarray}
  \fl 
K_\theta\left(\, x-y\ ;T\,\right) 
   &&=N\,
   \int  \left[\,Dp\,\right]\delta\left[\,\dot{\vec p} \,\right]
   \exp\left\{\, i\left[\, \vec p\cdot \vec x\,\right]_x^y -
  \int_0^T d\tau \left(\, \frac{\,\vec p\,{}^2}{2m} + \frac{\theta}{2T}\,\vec
  p\,{}^2\,\right)  \,\right\}\nonumber\\
  &&= \int \frac{d^2p}{(2\pi)^{2}}
   e^{i\vec p\cdot (\, \vec x-\vec y\,)} e^{-(\, T+m\theta\,)\,\vec p\,{}^2/2m }
   \end{eqnarray}
   
   The final form of the propagation kernel is 
   
   \begin{equation}
   K_\theta\left(\, x-y\ ;T\,\right)=\frac{1}{(2\,\pi)^2}\,\left(\, \frac{2\pi
   m}{T+m\theta}\,\right)
   \exp\left\{-\frac{m\,\left(\, \vec x-\vec y\,\right)^2}
   {2(\, T+m\theta\,)} \right\}
   \end{equation}

   Let us verify the short-time limit of the propagation kernel.
   The result is
   
   \begin{equation}
   K_\theta\left(\, x-y\ ; 0\,\right)=
   \left(\,\frac{1}{2\pi \,\theta}\,\right)
   \exp\left\{-\frac{\left(\, \vec x-\vec y\,\right)^2}
   { 2\theta} \, \right\}
   \end{equation}
   
   It shows that the kernel is not a delta-function,but a Gaussian. 
   This is due to the fact that the best possible localization of the particle, in non 
   commutative space, is       within a cell of area $\theta$.\\
   Knowing the kernel, we can calculate 
   the Green function. It is defined by the Laplace transform as

   \begin{eqnarray}
   G_\theta\left(\, x-y\ ; E\,\right)\equiv &&
   \int_0^\infty dT\, e^{-ET} \, K_\theta\left(\,x-y\ ; T\,\right)
   \nonumber\\
   &&= \int \frac{d^2p}{(2\pi)^2}  e^{i\vec p\cdot (\, \vec x-\vec
   y\,)}\,  G_\theta\left(\,E\ ; \,\vec p\,{}^2\,\right)
   \end{eqnarray}

   where the momentum space Green function is given by
   
   \begin{equation}
   G_\theta\left(\,E\ ; \,\vec p\,{}^2\,\right)= \left(\, \frac{1}{2\pi}\,\right)^2
   \, \frac{\exp\left(\, -\theta\, \vec p\,{}^2/2\,\right)}
   {E+ \frac{\,\vec p\,{}^2}{2m}}
   \end{equation}
   
   $G_\theta\left(\,E\ ; \,\vec p\,{}^2\,\right)$
   shows an \textit{exponential cut-off} for large momenta.\\ 
   
   The above calculation refers to a non-relativistic free particle and a relativistic 
   extension will be formulated in what follows. We are working
   in $2+1$ dimensions and keep, as usual, the (Euclidean) time as a commuting
   variable. We would like to mention the fact that the non commutative parameter
   $\theta$ selects preferred spatial directions leading to a violation of
   particle Lorentz invariance.  This fact has already been addressed in 
   non commutative models \cite{lorentz} and we shall just follow the accepted wisdom. The 
   relativistic version of $G_\theta\left(\, x-y\ ; E\,\right)$ is  
  
   \begin{equation}
  \fl 
G\left(\,  x-y\ ; m^2\,\right)\equiv N\int \left[\, De\,\right]
   \left[\, Dx\,\right] \left[\,
  Dp\,\right]\exp\left\{\, i\int_y^x  p_\mu\, d x^\mu -
  \int_0^T d\tau\left[\,
   e(\tau)\, \left(\, p^2 +m^2\,\right) + \frac{\theta}{2T}\,\vec p\,{}^2 
   \right] \,\right\}
   \end{equation}
   
   where, $e(\tau)$ is a Lagrange multiplier enforcing mass shell condition
   for the relativistic particle.
   
   \begin{equation}
   \fl 
G\left(\,  x-y\ ; m^2\,\right)= N\int \left[\, De\,\right]
   \int \frac{d^3p}{(2\pi)^3} e^{i p_\mu (\,  x- y\,)^\mu}
   \exp\left\{\, -\left(\, p^2 +m^2\,\right)
  \int_0^T d\tau e(\tau)-\theta\,\vec p\,{}^2/2\, \right\}
   \end{equation}
   
   The integration over $x(\tau)$ and $p(\tau)$ is carried out as in the 
   non relativistic case. The integration of the Lagrange multiplier $e(\tau)$ is   
   carried out by the introduction of the Feynman-Schwinger ``proper-time'' $s$ 
   
   \begin{equation}
   1=\int_0^\infty ds\,\delta\left[\, s  -\int_0^T d\tau e(\tau)\,\right]
   \end{equation}
   
   The final result for the relativistic case is
   
   \begin{eqnarray}
  \fl 
G_\theta\left(\, x-y\ ; m^2\,\right)&&= N\int_0^\infty ds\,
   e^{-s\, m^2}
   \int \frac{d^3p}{(2\pi)^3}\, e^{i p_\mu\, (\, x-y\,)^\mu}
   \exp\left[\, -(\, s+\theta\,) \,\vec p\,{}^2/2 \,\right]\nonumber\\
   &&\equiv \int\frac{d^3p}{(2\pi)^3}\, e^{i p_\mu \, (\, x- y\,)^\mu}
    G_\theta\left(\, p^2\ ; m^2\,\right)
   \end{eqnarray}
   
   where, the Feynman propagator in momentum space is given by
   
   \begin{equation}
   \fl 
G_\theta\left(\,\,\vec p\,{}^2\ ; m^2\,\right)=\left(\, \frac{1}{2\pi}\,\right)^3
   \frac{\exp\left[\, -\theta \,\vec p\,{}^2/2 \,\right]}{ p^2 +m^2}
   \end{equation}

   Thus, we arrive to the conclusion, already formulated in \cite{noilast},that the  non 
   commutativity of space leads to a an exponential cut-off in the Green function at large 
   momenta. The corresponding quantum field theory is UV finite since the loop diagrams e
   xhibit no divergences due to the non commutative cut-off.\\

\end{document}